\documentclass[doublecol, figures]{epl2}
\usepackage{amsmath}
\usepackage{amsfonts}
\usepackage{graphicx}
\usepackage{amssymb}
\usepackage{epsfig}

\def\lsim{\mathrel{\rlap{\lower4pt\hbox{\hskip1pt$\sim$}}
    \raise1pt\hbox{$<$}}}                
\def\gsim{\mathrel{\rlap{\lower4pt\hbox{\hskip1pt$\sim$}}
    \raise1pt\hbox{$>$}}}                

\begin{document}






\title{Benford's Law Detects Quantum Phase Transitions\\ similarly as Earthquakes}
\shorttitle{Benford's Law Detects Quantum Transitions similarly as Earthquakes}


\author{Aditi Sen(De), Ujjwal Sen}
\shortauthor{Aditi Sen(De) \etal}

\institute{Harish-Chandra Research Institute, Chhatnag Road, Jhunsi, Allahabad 211 019, India} 

\abstract{
A century ago, it was predicted that the first significant digit appearing in a data would be nonuniformly distributed, with the number one 
appearing with the highest frequency. This law goes by the name of Benford's law. 
It holds for data 
ranging from infectious disease cases to national greenhouse gas emissions. 
Quantum phase transitions are cooperative phenomena where qualitative changes occur in many-body systems at zero temperature.
We show that the century-old Benford's law can detect quantum 
phase transitions, much like it detects earthquakes. Therefore, being certainly of very different physical origins, seismic activity
and quantum cooperative phenomena may be detected by similar methods. 
The result has immediate implications in precise measurements in experiments in general, and for realizable quantum computers in particular. 
It shows that estimation of the first significant digit of measured physical observables 
is enough to detect the presence of quantum phase transitions in  macroscopic systems.
}

\maketitle

\section{Introduction}

Benford's law is an empirical law, first observed by Newcomb in 1881 \cite{Newcomb} and then 
by Benford in 1938 \cite{Benford}, predicting an uneven distribution of the 
digits one through nine at the first significant place, for data obtained in a huge variety of situations. Precisely, the prediction is that the 
frequency of occurence of the digit \(D\) will be 
\begin{equation}
\label{kyano-tomra-amai-daako-Biswarup-dar-mobile-er-ringtone}
P_D= \log_{10}\left( 1+ \frac{1}{D}\right).
\end{equation}
This implies a much higher occurence of one -- about 30\% of the cases -- in comparison to the higher digits.
The higher digits are predicted to appear with progressively lower frequencies of occurence, with e.g. nine appearing with about 5\% probability.

The law has since been checked for a wide spectrum of situations in the natural sciences, as well as for mathematical series \cite{website}.
The situations in the natural sciences range from the number of cases of infectious diseases occuring globally, to 
the national greenhouse gas emissions \cite{infect, green, Sambridge, website}. 
Mathematical insights into the Benford's law were recently obtained in a series of papers by Hill \cite{Hill, Hill1, Hill2}.

Interestingly, it was recently discovered by Sambridge \emph{et al.} \cite{Sambridge} that
the Benford's law could be used to distinguish earthquakes from background noise, and the thesis was successfully applied to seismographic data from  
the Boxing Day Sumatra-Andaman earthquake in 2004.

We find 
that the Benford's law can be used to detect the position of a (zero temperature) quantum phase transition in a quantum many-body system.
A phase transition in a bulk system is a qualitative change in one or more physical  quantities characterizing the system, and their 
importance appear in diverse natural phenomena.
 Phase transitions could 
be temperature-driven, like the transformation of ice into water, and are produced by thermal 
fluctuations \cite{classical-Sachdev}. \emph{Quantum} phase transitions, however, occur at zero 
temperature, and are driven by a system parameter (like, magnetic field), and they ride on purely quantum fluctuations \cite{Sachdev}. Their importance can hardly be
overestimated, and range from fundamental aspects -- like understanding the appearance of pure quantumness in bulk matter 
-- to 
revolutionizing applications -- like using the Mott insulator to superfluid transition  for realizing quantum computers \cite{MI-to-SF, Bloch08, QC, Lewenstein07}.

We consider a paradigmatic quantum many-body system, the quantum transverse Ising model, which exhibits a quantum phase transition \cite{eita-prothhom, TIM, Sachdev}. 
There are solid state compounds that can be described well by this model. In particular, the
\(\mbox{Li}\mbox{Ho}_x\mbox{Y}_{1-x}\mbox{F}_4\) compound is known to be described by the 
three-dimensional (quantum) transverse  Ising
model
\cite{solid-ladder}.
Moreover, spectacular recent advances in cold gas experimental 
techniques have made this  model realizable in the laboratory, with the additional feature that 
dynamics of the system can be simulated by controlling the system parameters and the applied transverse field. 
In particular, the two-component Bose-Bose and Fermi-Fermi mixture, in the strong coupling limit with suitable tuning of scattering length and additional 
tunneling in the system can be described by the quantum \(XY\) Hamiltonian, of which the Ising Hamiltonian is a special case \cite{Lewenstein07, Bloch08, iontrap}. 
Therefore the phenomenon discussed in this paper 
can be verified in the laboratory with curently available technology. Let us note here that the quantum phase transition in the transverse Ising model has been 
experimentally observed \cite{Ising-experiment}.

\section{The Hamiltonian governing the system, and its diagonalization}

 The quantum transverse \(XY\) system is described on a lattice by the Hamiltonian
\begin{equation}
\label{Snajh-ghumoch-chhe}
H = J \sum_{\langle ij \rangle} \left[(1 + \gamma) S_i^x S_{j}^x + (1 - \gamma) S_i^y S_{j}^y\right] - 
a \sum_i S_i^z,
\end{equation}
where \(J\) denotes the coupling constant, \(\gamma\) the anisotropy parameter, and \(a\) is the transverse field strength. 
\(S_i^x\), \(S_i^y\), \(S_i^z\) are one-half of the Pauli matrices \(\sigma_i^x\), \(\sigma_i^y\), \(\sigma_i^z\) respectively
  at the \(i\)th site. \(\langle ij \rangle\) indicates that the corresponding summation
runs over nearest neighbor spins on the lattice. For \(\gamma=1\), the system reduces to the quantum transverse Ising model. 
For our purposes, we will consider the model on an infinite  one-dimensional lattice, and in this case, the Hamiltonian is exactly diagonalizable 
by successive Jordan-Wigner, Fourier, and Bogoliubov transformations \cite{LSM61, Barouch1075, Barouch786}.

We now diagonalize the system and find the single-site as well as two-site physical quantities for the ground state (at zero temperature)
\cite{LSM61, Barouch1075, Barouch786}. 
Let us denote the ground state by \(\varrho\). The single-site state is described by a single physical quantity, the transverse magnetization:
\begin{equation}
M^z = \lim_{n\to \infty}\mbox{tr}\left(\frac{1}{n}\sum_i\sigma_i^z \varrho\right).
\end{equation}
The two-site state additionally has the three diagonal correlations:
\begin{equation}
C^{\alpha \alpha} = \lim_{n\to \infty}\mbox{tr}\left(\frac{1}{n}\sum_i\sigma_i^\alpha \sigma_{i+1}^\alpha \varrho\right),
\end{equation}
\(\alpha = x, y, z\).
(Periodic boundary condition is assumed here.) The magnetization and nearest neighbor correlations are given as follows. 
\begin{equation}
C^{xx}(\tilde{a})=G(-1, \tilde{a}), \quad 
C^{yy}(\tilde{a})=G(1, \tilde{a}),
\end{equation}
and
\begin{equation}
C^{zz}(\tilde{a})=  [M^z(\tilde{a})]^2 - G(1, \tilde{a})G(-1, \tilde{a})
\end{equation}
where 
$G(R, \tilde{a})$ (for $R =\pm 1$) is given by
%
%
%
\begin{equation}
G(R, \tilde{a}) =  \frac{1}{\pi}\int^\pi_0d\phi \frac{1} {\Lambda(\tilde{a})} 
 (\gamma \sin(\phi R)\sin \phi - \cos\phi (\cos\phi -\tilde{a})) 
\end{equation}
And 
\begin{equation}
M^z(\tilde{a}) = -\frac{1}{\pi} \int_0^\pi d\phi \frac{ (\cos\phi - \tilde{a})}{\Lambda(\tilde{a})} 
\end{equation}
Here 
\begin{equation}
\Lambda(x)= \left\{\gamma^2\sin^2\phi~+~[x-\cos\phi]^2\right\}^{\frac{1}{2}},
\end{equation}
and 
\begin{equation}
 \tilde{a} = \frac{a}{J}. 
\end{equation}
Note that \(\tilde{a}\) is a 
dimensionless variable, and
in the following, we will use it as the field parameter. 

\section{A measure of entanglement} 

Apart from the classical correlations, \(C^{\alpha \alpha}\), and the transverse magnetization, the two-site state also possesses
quantum correlations aka entanglement \cite{Horodecki09}, which is increasingly being used to describe and characterize phenomena in 
many-body systems \cite{Wootters02, Osterloh02, Osborne02, Lewenstein07, Amico08}. 
There are several measures of quantum entanglement, and here we choose the logarithmic negativity \cite{Vidal-Werner} as a measure 
for the two-site state under consideration. It is defined for a two-site state \(\rho_{AB}\) as 
\begin{equation}
  E_{N}(\rho_{AB}) = \log_2 [2 {\mathcal{N}}(\rho_{AB}) + 1],
\label{raat-duto-bajey-kaal-khajuraho-jabo}
\end{equation}
where the negativity \({\mathcal{N}}(\rho_{AB})\)
is defined as the absolute value of the sum of the negative
eigenvalues of \(\rho_{AB}^{T_{A}}\), with \(\rho_{AB}^{T_{A}}\) being the partial transpose of \(\rho_{AB}\) with respect to the \(A\)-part \cite{Peres_Horodecki}.

\section{The Benford quantity and the violation parameter}

For checking the status of the Benford's law for a given quantity \(Q\), it is necessary to suitably shift and scale it. To see the necessity of this exercise, 
consider the situation where we want to check the status of the Benford's law for the voltmeter reading of a particular electrical circuit. Suppose that the 
power provider promises that the reading will be 230V, with usual small deviations. Let us assume that the deviation is never more than 10V, so that the voltmeter reading 
will always be between 220V and 240V. No matter how many readings we take, the first significant digit will always be 2, a clear, but trivial, violation of 
 Benford's law. 
%
%
A simple way to get around this problem is to shift and scale 
the quantity, so as to bring it to the range \((0,1)\) \cite{aaj-s(n)ajh-er-jor-elo-abar}.

Therefore, before checking the status of the Benford's law for a particular physical quantity \(Q\), 
of the quantum spin model under consideration, we first shift its origin and scale it, so as to bring it to the range
\((0,1)\) as follows:
\begin{equation}
\label{prochur-mosha}
Q_B = \frac{Q - Q_{\min}}{Q_{\max} - Q_{\min}}.
\end{equation}
We then check the validity of the Benford's law for the ``Benford \(Q\)'', \(Q_B\). 
Here \(Q_{\min}\) and \(Q_{\max}\) are respectively the minimum and maximum of the 
physical quantity \(Q\), in the relevant range of operation.

As a measure to quantify the amount of potential violation of the Benford's law for \(Q\), we consider the quantity 
\begin{equation}
\delta (Q_B) = \sum_{D=1}^9 \left| \frac{O_D - E_D}{E_D}\right|.
\end{equation}
Here \(O_D\) is the observed frequency of the digit \(D\) as the first significant digit of \(Q_B\), chosen from the sample under consideration. 
\(E_D\) is the expected frequency for the same, so that 
\begin{equation}
\label{snajh-ghomolo-ki}
E_D = N P_D,
\end{equation}
 where \(N\) is the sample size, and \(P_D\) is the probability expected from the Benford's law
(see Eq. (\ref{kyano-tomra-amai-daako-Biswarup-dar-mobile-er-ringtone})).

Note that the shifting and scaling process mentioned above produces a zero and unit value for \(Q_B\), which are then removed from the data set, as trivial 
data points. Correspondingly the sample size is also reduced by \(2\), and this reduced value is named \(N\), and used in Eq. (\ref{snajh-ghomolo-ki}).

\vspace{0.7cm}
\begin{figure}[h]
\label{fig-chhobi-prothhom}
\begin{center}
\epsfig{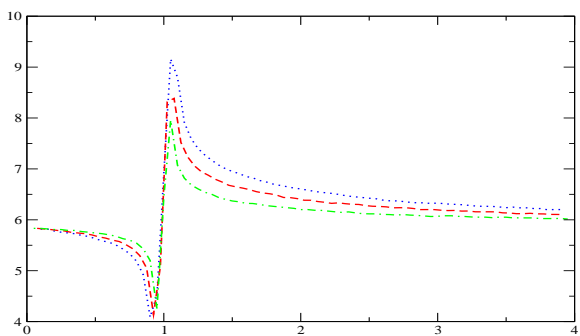}
\caption{(Color online.) Benford's law detects the quantum phase transition in the infinite transverse Ising model. 
The horizontal axis represents the physical parameter \(a/J\), 
while the vertical one represents the violation parameter for transverse magnetization. Both axes depict dimensionless quantities.
The blue dotted, red dashed, and green dot-dashed curves are respectively for shifting field windows (or more precisely for shifting \(a/J\) windows) of 
lengths 0.2, 0.15, and 0.1. Each of the curves have been drawn for \(N=1998\), but is equal to the same for \(N=1498\), indicating that convergence 
with respect to \(N\) has already been attained. For the plots, the shifting field window is assumed to shift with discrete jumps of \(0.05\).
%
}
\end{center}
\end{figure}

\section{Status of the Benford's law on a shifting field window}

Let us begin by considering the status of the Benford's law for the transverse magnetization in the transverse Ising model. We wish to scan the 
status of the law as we move along the axis of the transverse field \(a\). For a given value of the applied transverse field \(a=a{'}\), we choose 
a \emph{field window} 
\begin{equation}
(a{'}/J - \epsilon/2, a{'}/J + \epsilon/2),
\end{equation}
 for  small \(\epsilon\), and choose \(N\) points from this field window. We then find 
the transverse magnetization of the system for these \(N\) values of the transverse field. 
We shift and scale these values of the magnetization to find the Benford transverse magnetization (see Eq. (\ref{prochur-mosha})).
 These \(N\) values of the Benford transverse magnetization form 
the sample for this field window to which we fit the Benford's law. We find the corresponding \(O_D\)'s, and then the corresponding \(\delta\) 
\cite{accuracy}. 
The lower the value of the \(\delta\) is, the more is the Benford's law satisfied in that window. We scan the field axis, finding the values of the 
\(\delta (M^z_B)\) as we move. We find that the Benford's law is more violated in the quantum paramagnetic 
region (\(a/J >1\)), than in the 
magnetically ordered one
(\(0<a/J<1\)).

\begin{figure}[h]
\includegraphics[width=1.7 in, height=1.3 in]{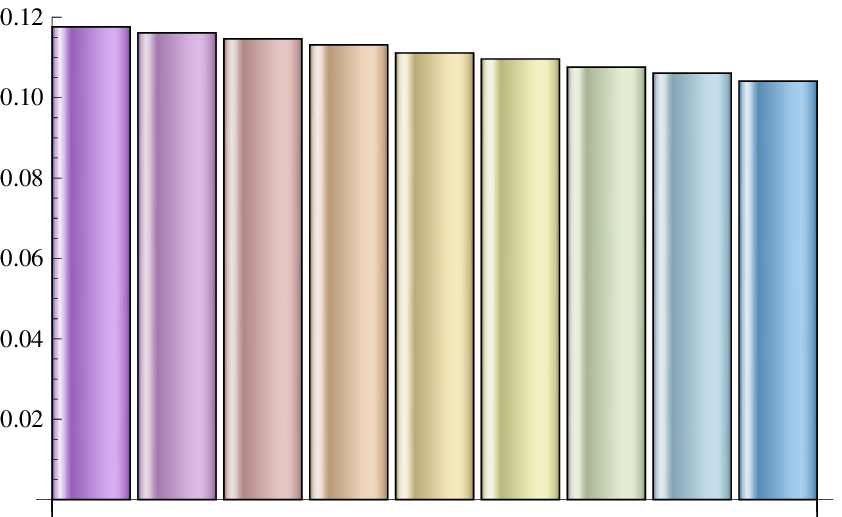}
\includegraphics[width=1.7 in, height=1.3 in]{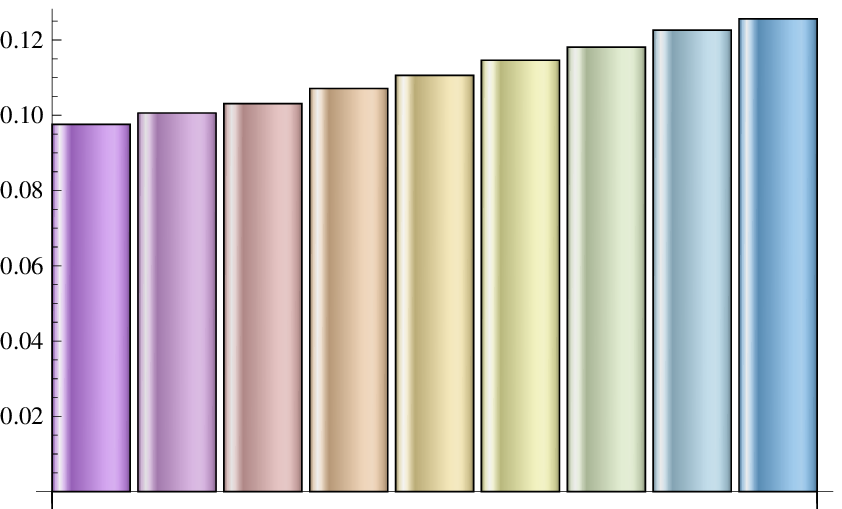}
\caption{(Color online.) A comparison of the different behavior of the relative frequencies of the first significant digits of the transverse magnetization of the 
infinite transverse 
Ising model, before and after the quantum phase transition in the model. As stated in the text, 
the frequencies are actually for the Benford transverse magnetization (see Eq. (\ref{prochur-mosha})). 
 The histogram on the left is for the field window \(a/J \in (0.82,0.9)\), while that 
on the right is for \(a/J \in (1.1,1.18)\). The number of data points is 1998 for both the field windows. For each histogram, 
the nine columns, from left to right, are for the nine first significant digits 1 through 9. 
Each column height is the relative frequency for the corresponding significant digit. 
A clearly different behavior is seen in the relative frequency distributions
before and after the quantum phase transition. Both axes of both the histograms represent dimensionless quantities.}
\label{chhobi-tritiyo}
\end{figure}

However, the most interesting result is that \(\delta (M^z_B)\) clearly signals the position of the quantum phase transition at \(a/J=1\). 
Away from the quantum phase transition, the violation parameter \(\delta (M^z_B)\) seems to have ``equilibriated'', and is more or less a constant as we 
move the field window over the \(a\) axis. However, at the quantum phase transition, there is a sudden and vigorous transverse movement of the 
violation parameter. See Fig. 1. The constant values of 
the violation parameter before and after the vicinity of the quantum phase transition at \(a/J=1\) are different, and the transverse movements are  
created as the system tries to change its equilibriation value of the violation parameter. 
While the amount of the transverse movement and the length 
of the field window that is required 
for the violation parameter to equilibriate, depend on the length of the field window, all field 
windows lead to the same point on the field axis at which the transverse movement occurs.
The situation is very similar to the detection of the Sumatra-Andaman earthquake by Sambridge \emph{et al.} \cite{Sambridge},
where there appeared a distinct change in the violation parameter considered by them, when the earthquake entered their shifting \emph{time window},
which was otherwise scanning over a background noise. 
Surprisingly therefore, although the quantum phase transition of the transverse Ising model is of a completely different origin as compared to  the 
seismic activities below the surface of the earth, the Benford's law detects them very similarly.

The violation parameter is a characteristic of the frequency distribution of the first significant figure corresponding to the 
physical quantity under consideration. The distribution itself hides further information, and in particular can also be directly used to 
detect the quantum phase transition. And for instance, the behaviors of the relative frequency distribution for the transverse magnetization, before and after the transition, 
are significantly different.  See Fig. 2.


\vspace{0.4cm} 
\begin{figure}[h!]
\label{fig-chhobi-dwitiyo}
\begin{center}
\epsfig{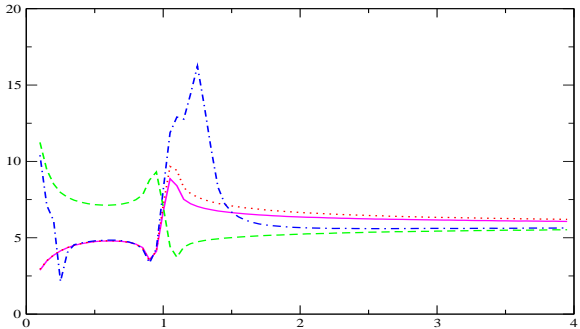}
\caption{(Color online.) Comparing the violation parameters for different physical parameters of the infinite transverse Ising model.  
The horizontal axis represents the physical parameter \(a/J\). 
The vertical one represents the violation parameter for the Benford \(Q\), where \(Q\) denotes the physical parameters 
\(C^{xx}\) (nearest neighbor \(xx\) classical correlation, plotted as a pink continuous curve), 
\(C^{zz}\) (nearest neighbor \(zz\) classical correlation, plotted as a red dotted curve),
single-site von Neumann entropy (green dashed curve), 
and  
entanglement (logarithmic negativity, plotted as a blue dot-dashed curve). 
Both axes represent dimensionless quantities.
The shifting field windows are of length 0.2, with the shifts being of length 0.05, for all the curves. 
The plots are for 3998 sample points. Convergence 
with respect to the number of sample points was checked by using \(N= 1498\), 1998, and 2998. Note that the curves for \(C^{xx}\) and \(C^{zz}\)  almost 
coincide for \(a/J < 1\).
}
\end{center}
\end{figure}

\begin{figure}[h!]
\label{fig-chhobi-chaturthho}
\begin{center}
\epsfig{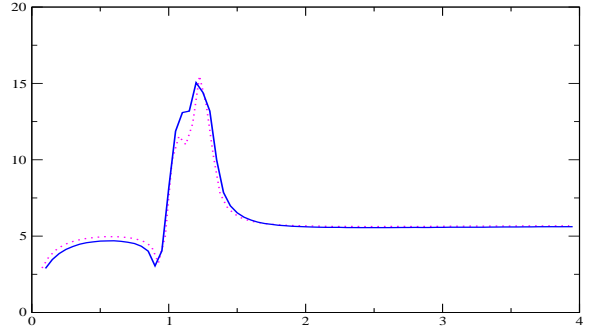}
\caption{(Color online.) Bounding equilibriated plateaus of different heights that are robust under 
small changes of the length of the shifting field window are necessary to detect phase transitions. 
The Benford nearest neighbor \(yy\) classical correlation for the infinite transverse Ising model is plotted on the vertical axis against 
the field parameter \(a/J\) on the horizontal axis.
Both axes represent dimensionless quantities.
The shifting field window is of length 0.15 for the pink dotted curve, and is 0.2 for the blue continuous curve. 
The shifts are of length 0.05 for both the curves. The plots are for 2998 sample points. Convergence 
with respect to the number of sample points was checked by using \(N= 1498\) and 1998. 
%
}
\end{center}
\end{figure}


Phase transitions, and in particular, quantum phase transitions can be detected by a variety of indicators, including (but certainly not limited to)
classical correlations \cite{Sachdev}, bipartite and multipartite quantum entanglement \cite{Wootters02, Osterloh02, Osborne02, Lewenstein07, Amico08, multi-qpt}, 
quantum discord \cite{discord-qpt}, ground state fidelity and fidelity susceptibility \cite{fid-qpt}, etc.
The quantum phase transition in the transverse Ising model can 
also be detected by using the violation parameter for other physical quantities, like the classical correlations, and
the quantum correlation, \(E_N\). Quantum phase transitions are present also in the quantum \(XY\) models with other values of the anisotropy \(\gamma\), and 
these can also be detected by using the violation parameter for the different physical quantities in those models.
In Fig. 3, we plot the violation parameter for the Benford \(Q\) (see Eq. (\ref{prochur-mosha})), where \(Q\) denotes the physical parameters \(C^{xx}\)
(nearest neighbor \(xx\) classical correlation), \(C^{zz}\) (nearest neighbor \(zz\) classical correlation), 
the single-site von Neumann entropy \cite{vonN}, and logarithmic negativity, and again find that 
the quantum phase transition at \(a/J=1\) is detected by bi-directional (i.e. both downwards and upwards)
 transverse movements of the violation parameter, bounded on both sides by equilibriated 
plateaus of different heights. A bi-directional transverse movement is also seen at \(a/J=0\) in
 the violation parameter for the Benford logarithmic negativity. However, it is not accompanied by bounding equilibriated plateaus of 
different heights. A small bi-directional transverse movement is seen in the violation parameter for the Benford \(C^{yy}\) (Fig. 4) at approximately
\(a/J =1.1\).  However, again it is not bounded by equilibriated plateaus, and moreover, it vanishes for a slight change of the length of the 
shifting field window. [A similar feature was also observed for the Benford \(E_N\), for 
\(\epsilon = 0.15\), which got erased for \(\epsilon =0.2\).] 
Let us also note here that neither a maximum nor a minimum of the violation parameter indicates a quantum phase transition -- 
 maxima (minima) of the violation parameter are obtained for the single-site entropy and entanglement (\(C^{xx}\), \(C^{yy}\), and \(C^{zz}\)) at \(a/J=0\).
These transverse movements at \(a/J=0\) are however not of the typical ``N''-shape of a transverse motion (as for the ones at \(a/J=1\)), and 
moreover are bounded by equilibriated plateaus of \emph{equal} heights. 


There are two interesting differences between 
the violation parameters for the different physical quantities. 
Firstly, the violation parameter for magnetization has the largest delay in returning to equilibrium in comparison to the situation for the other parameters.
Secondly, the transverse movement of the violation parameter at the quantum phase transition for entanglement and \(C^{yy}\) 
is more than three times 
that for 
magnetization and other parameters.
It is interesting to compare these differences, and in particular the similarity of the behavior of 
entanglement with a classical correlation, with the usual 
fragile nature of entanglement.  See e.g. \cite{ref-JJ, Haroche, atomsexp, nmrexp, Bloch08, iontrap, Zeilinger}. Also, the more violent 
transverse motion for entanglement and \(C^{yy}\) seem to imply that the violation parameter for them 
can act as better detectors 
of quantum phase transitions.
Comparing the behaviors of the violation parameters of the different physical quantities, it seems plausible 
that the quantum phase transition is signaled by
the maximum or minimum value (whichever is of higher modulus) of the derivative of the corresponding violation parameter.

Let us mention here  that 
the violation parameter for all physical quantities  considered except the single-site entropy,
is higher in the paramagnetic region than in the magnetically ordered one.

\vspace{0.4cm} 
\begin{figure}[h!]
\label{fig-chhobi-pancham}
\begin{center}
\epsfig{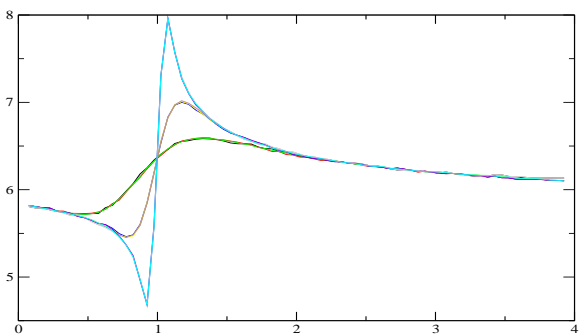}
\caption{(Color online.) The transition is already visible in the violation parameter of the Benford's law for 
finite systems. 
The Benford transverse magnetization for finite transverse Ising models is plotted on the vertical axis against 
the field parameter \(a/J\) on the horizontal axis. 
Both axes represent dimensionless quantities.
The shifting field window is of length 0.15, and 
the shifts are of length 0.05 for all the curves. 
The innermost curve represents the violation parameter for a system of 10 spins, and the outermost curve is that of 100 spins, while 
the intermediate one is of 25 spins. Each of the curves is actually an amalgamation of three curves obtained by taking respectively 
1498, 1998, and 2998 sample points, and they almost merge with each other for a particular value of the total number of spins. 
%
}
\end{center}
\end{figure}

An appealing feature of the transverse \(XY\) model is that it can be realized in different physical systems. In 
particular, finite chains of spins described by this model Hamiltonian can be realized with ultracold gases \cite{Lewenstein07, Bloch08, iontrap}.
 In the light of this, it is interesting 
to know whether the violation parameter for different physical quantities can show traces of the quantum phase transition 
already in finite chains. For a chain of \(n\) spins, the transverse magnetization is given by \cite{LSM61, Barouch1075, Barouch786} 
\begin{equation}
M_n^z(\tilde{a}) = - \frac{1}{n}\sum_{p=1}^{n/2}\frac{\cos\phi_p -\tilde{a}}{\Lambda_p(\tilde{a})},
\end{equation}
where 
\begin{equation}
 \Lambda_p(\tilde{a}) = \left\{\gamma^2\sin^2\phi_p~+~[x-\cos\phi_p]^2\right\}^{\frac{1}{2}},
\end{equation}
with 
\begin{equation}
 \phi_p = \frac{2 \pi p}{n}.
\end{equation}
Here we have assumed the so-called ``c-cyclic'' boundary conditions (see Refs. \cite{LSM61, Barouch1075}). 
In Fig. 5, we consider the violation parameter for the Benford transverse magnetization for finite chains of different lengths, and
show that traces of the quantum phase transition are already distinctly visible in such systems.

\section{Conclusion and discussions}

A century old empirical law, known as Benford's law, predicts that the first significant digit in data, obtained either from natural phenomena or 
from mathematical tables, will be nonuniformly distributed, with decreasing frequency of occurrence of the digits one through nine. 
We have shown that the Benford's law can be used to detect cooperative quantum phenomena in many-body systems. Specifically, we have shown that 
the (zero temperature) quantum phase transition in the quantum transverse \(XY\) model of spin-1/2 particles 
can be detected by the amounts of violation  of the Benford's law by 
several physical quantities  of the spin system. Interestingly, the nature of the detection of the quantum phase transition is similar 
to that of a recently discovered method of detecting earthquakes in a seismic data containing both background noise and earthquake information of the 2004 Boxing Day 
earthquake in the Sumatra-Andaman region of the Indian ocean.

Apart from its fundamental implications, the first application-oriented 
implication of the result is the following. Although seismic activity and quantum phase transitions have very different physical origins,
they are both detected by the Benford's law in quite a similar method. It may therefore be possible that other methods 
known in one of these streams of study can be successfully applied to the other stream. 
In particular, this may help us to identify new ways to tackle the problems at the interface of quantum information science and condensed matter physics.

Secondly, and more directly, 
the results obtained here have immediate applications in  actual experimental detection of quantum phase transitions.
%
%
The result shows that the first significant digit is enough to detect a quantum phase transition in a many-body system realized in the laboratory, 
and that quantum fluctuations in a physically realizable system can already be detected by looking at the first significant digit. 
Decoherence and other noise mechanisms remain a pertinent problem in dealing with quantum many-body  systems, 
and have e.g. been a reason for scalability problems of quantum computing devices 
(see e.g. Refs. \cite{ekta-ponchanno-baje}, and references therein). 
The realization that quantum phase transitions can be detected by observing the first significant digits of physical quantities may open up new ways of 
handling noise effects in these systems.
The result
gives us further 
reasons to believe that the battle 
for controlling quantum many-body systems, including that
against the current limitations of quantum information processing in many-body systems, can be won.


\begin{thebibliography}{999}
\bibitem{aaj-s(n)ajh-er-jor-elo-abar} Another possibility could be 
 to take the deviation around the
mean reading. However, that will result in an artificial approximate double counting. To avoid the latter, 
we may consider only positive or negative deviations, which however will be unnatural if the distribution is skewed about the mean. Samples obtained by considering deviations about the mean would also be unnatural if the kurtosis were high.
\bibitem{accuracy} In numerically calculating \(O_D\),We ignore sample points whose first significant digit is at the sixth 
or higher decimal places, in the Benford observables 
(see Eq. (\ref{prochur-mosha})).
\bibitem{vonN} The von Neumann entropy of a quantum state \(\varrho\) is defined as \(-\mbox{tr}(\varrho \log_2 \varrho)\).
\end{thebibliography}
\end{document}